\documentclass[conference]{IEEEtran}
\usepackage{cite}
\usepackage{amsmath,amssymb,amsfonts,bm}
\usepackage[ruled,vlined]{algorithm2e}
\usepackage{graphicx}
\usepackage{subcaption}
\usepackage{textcomp}
\usepackage[table]{xcolor}
\usepackage{booktabs}
\usepackage{threeparttable}
\usepackage{multirow}
\usepackage{array}
\usepackage{colortbl}
\usepackage[colorlinks,urlcolor=blue,linkcolor=blue,citecolor=blue]{hyperref}
\usepackage{listings}

\lstdefinestyle{ag}{
basicstyle=\ttfamily\scriptsize,
breaklines=true,
columns=fullflexible,
frame=single,
framesep=3pt,
backgroundcolor=\color{black!3},
commentstyle=\color{teal!70!black}
}

\definecolor{winrow}{rgb}{0.85,0.92,1.0}
\definecolor{recallc}{rgb}{0.80,0.95,0.80}
\definecolor{qualc}{rgb}{1.0,0.90,0.78}

\begin{document}

\title{Agentra: A Supervisable Multi-Agent Framework for Enterprise Intrusion Response}

\DeclareRobustCommand{\IEEEauthorrefmark}[1]{\smash{\textsuperscript{\footnotesize #1}}}

%\author{Anonymous Author}
\author{
    \IEEEauthorblockN{
        Raj Patel\IEEEauthorrefmark{1}\IEEEauthorrefmark{3},
        Shaswata Mitra\IEEEauthorrefmark{1}\IEEEauthorrefmark{4},
        Michele Guida\IEEEauthorrefmark{2}\IEEEauthorrefmark{5},
        Stefano Iannucci\IEEEauthorrefmark{2}\IEEEauthorrefmark{6},
        Sudip Mittal\IEEEauthorrefmark{1}\IEEEauthorrefmark{7},
        Shahram Rahimi\IEEEauthorrefmark{1}\IEEEauthorrefmark{8}
    }
    \IEEEauthorblockA{\IEEEauthorrefmark{1}The University of Alabama, Alabama, USA}
    \IEEEauthorblockA{\IEEEauthorrefmark{2}Roma Tre University, Rome, Italy}
    \{rpatel38\IEEEauthorrefmark{3},
        smitra3\IEEEauthorrefmark{4}\}@ua.edu,
        mic.guida1@stud.uniroma3.it\IEEEauthorrefmark{5},
        stefano.iannucci@uniroma3.it\IEEEauthorrefmark{6},
        \{smittal1\IEEEauthorrefmark{7},
            srahimi1\IEEEauthorrefmark{8}\}@ua.edu
}

\maketitle

\begin{abstract}
Enterprise intrusion response still depends on static playbooks and analyst-driven triage, creating delay between alert generation and containment. We present Agentra, a supervisable multi-agent Intrusion Response System (IRS) framework that converts alerts from IDS, EDR, and XDR platforms into structured incident response plans grounded in MITRE ATT\&CK, MITRE D3FEND, and NIST CSF 2.0. Agentra decomposes response reasoning across role-scoped agents, validates proposed plans through a bounded Planner--Validator review loop, screens retrieved threat intelligence through a Moderator security gateway, gates actions through an Action Catalog and risk score, and records decisions in an append-only audit log. We evaluate Agentra against a static OASIS CACAO v2.0 cyber-playbook baseline on a 120-event corpus drawn from ThreatHunter-Playbook, Splunk BOTSv3, and DARPA OpTC. The strongest configuration improves FP-aware IRS $F_{1}$ from $0.61$ to $0.84$ and restores the projected harmful-action rate to the static baseline level of 0.0\% after Planner-only configurations introduce unsafe overreaction. These results indicate that multi-agent response planning can improve ontology-grounded IRS coverage while preserving analyst approval and auditability.
\end{abstract}


\section{Introduction} \label{introduction}
Enterprise security platforms can surface incidents quickly, but response remains constrained by static playbooks and analyst-driven triage. CrowdStrike reports an average eCrime breakout time of 29 minutes in 2025, with the fastest observed breakout at 27 seconds, while Unit 42 reports that the fastest quartile of intrusions reaches exfiltration in 1.2 hours~\cite{crowdstrike2026,unit42_2026}. At the same time, IBM reports a mean breach lifecycle of 241 days and an average United States breach cost of \$10.22 million~\cite{ibm2025costbreach}. This gap shows that detection has accelerated faster than response.

Modern IDS, EDR, and XDR platforms aggregate endpoint, identity, network, and cloud telemetry, yet response workflows still rely on Security Orchestration, Automation, and Response (SOAR) playbooks encoded as deterministic, analyst-maintained workflows~\cite{islam2019multivocalSOARreview}. These workflows support repeatable tasks but do not adapt well when adversaries vary tactics, chain actions across surfaces, or exploit gaps between siloed tools. Multi-vocal reviews of security orchestration also document recurring limits of workflow-centric automation, including rule maintenance burden and weak adaptation to previously unseen activity~\cite{islam2019multivocalSOARreview}. KuppingerCole identifies human cognitive capacity as a primary SOC constraint, and Unit 42 reports that many intrusions span multiple attack surfaces and require investigation across several telemetry sources~\cite{kuppingercole2026soc,unit42_2026}.

Recent work applies large language models to incident-response reasoning. IRCopilot automates incident response through LLM-driven analysis~\cite{lin2025IRCopilotLLM}, and Tellache et al. couple LLMs with cyber threat intelligence for autonomous response~\cite{tellache2025autonomousIRwithCTI}. These systems show that language models can interpret heterogeneous evidence and propose response actions, but enterprise execution also requires layered supervision, including independent review of proposed plans, mediated retrieval, bounded action scope, analyst approval, and an integrity-protected record of decisions.

The need for supervision is reinforced by work on LLM-integrated and tool-using agents. Greshake et al.~\cite{greshake2023indirectInjection} show that retrieved content can carry indirect prompt-injection payloads that alter model behavior, and AgentDojo evaluates injection attacks and defenses for agents that invoke external tools~\cite{debenedetti2024agentdojo}. This work indicates that, for an enterprise IRS, both retrieved threat intelligence and model-generated actions should be treated as untrusted rather than authoritative.

An agentic IRS is therefore needed because response planning requires more than selecting a stored playbook. A response system must interpret heterogeneous evidence, retrieve relevant internal and external context, map events to security ontologies, propose containment or remediation steps, and check whether those steps are safe for the affected environment. This paper presents Agentra (\textit{Agentic Response Action}), a supervisable multi-agent framework for enterprise intrusion response. Agentra builds on the structured, ontology-grounded character of playbook-based response while replacing static lookup with LLM-assisted planning constrained by supervision. Rather than pursue direct autonomous execution, Agentra decomposes response reasoning across agents and gates the resulting plan through retrieval moderation, Validator review, analyst approval, bounded actions, risk scoring, and audit logging.

We evaluate two questions. \textbf{(RQ1)} \textit{Can a supervisable agentic IRS extend static CACAO-based response planning while preserving contextual grounding and policy-aware planning?} \textbf{(RQ2)} \textit{What is the marginal contribution of each architectural component, and at what operational-safety cost?}

This paper makes the following contributions:
\begin{itemize}
\item Agentra, an agentic IRS framework that turns IDS, EDR, and XDR alerts into CACAO~v2.0 plans with traceable MITRE D3FEND mitigation tags.
\item A supervisable orchestration pipeline in which Planner-generated Intrusion Response Plans (IRPs) undergo bounded Validator review, analyst approval, Action Catalog screening, sanity checking, and risk scoring.
\item A Moderator-based Security Gateway that screens enterprise-local and external retrievals for allow-list violations, prompt-injection patterns, and out-of-ontology hallucinated mitigation identifiers.
\item A five-step ablation on a 120-event corpus drawn from ThreatHunter-Playbook, Splunk BOTSv3, and DARPA OpTC.
\end{itemize}

%SThe remainder of the paper is organized as follows. Section~\ref{sec:framework} describes the Agentra framework. Section~\ref{sec:experiments} reports the empirical evaluation. Section~\ref{sec:discussion} discusses limitations, and Section~\ref{sec:conclusion} concludes the paper.

\section{The Agentra Framework} \label{sec:framework}
Agentra is a response-side framework that converts alert evidence from existing security platforms into supervisable, ontology-grounded incident response plans. It consumes alert evidence but does not perform primary detection. This section defines the threat model, architecture, orchestration pipeline, and response controls.

\subsection{Threat Model} \label{sec:threat}
We assume deployed IDS, EDR, XDR, enterprise enforcement APIs, identity infrastructure, endpoint controls, and the underlying enterprise network are not compromised. The adversary is scoped to the agentic response layer. Within this layer, the adversary may introduce misleading, poisoned, or instruction-like content into artifacts retrieved or summarized by LLM agents, including external threat intelligence, incident notes, logs, or contextual documents exposed to retrieval. The adversary's goal is to cause the Planner, Moderator, or Validators to suppress containment, recommend an unsafe action, over-broaden the response, or operate outside its assigned role.

Agentra's objective is to prevent a single injected prompt, poisoned retrieval, or LLM-agent failure, including unsupported or hallucinated reasoning, from producing an unapproved or unsafe response. The framework treats retrieved content and model outputs as untrusted and constrains them through moderated retrieval, role-scoped agents, Validator review, bounded action catalogs, risk scoring, analyst approval, state-change checks, and append-only audit logging. These controls support supervisable response planning; they do not provide formal verification of LLM behavior.

These assumptions translate into six design goals: \textbf{G1}, generate high-quality response plans; \textbf{G2}, prevent autonomous actions from reaching production surfaces without analyst authorization; \textbf{G3}, preserve an integrity-protected record of prompts, retrievals, Validator votes, risk decisions, and analyst approvals; \textbf{G4}, ground classification and response in MITRE ATT\&CK, MITRE D3FEND, and NIST CSF~2.0~\cite{nist_csf2_2024}; \textbf{G5}, prevent LLM-side vulnerabilities introduced by agentic orchestration; and \textbf{G6}, prevent approved plans from executing against materially changed system state without revalidation.

\begin{figure*}[!ht]
\centering
\includegraphics[scale=0.26]{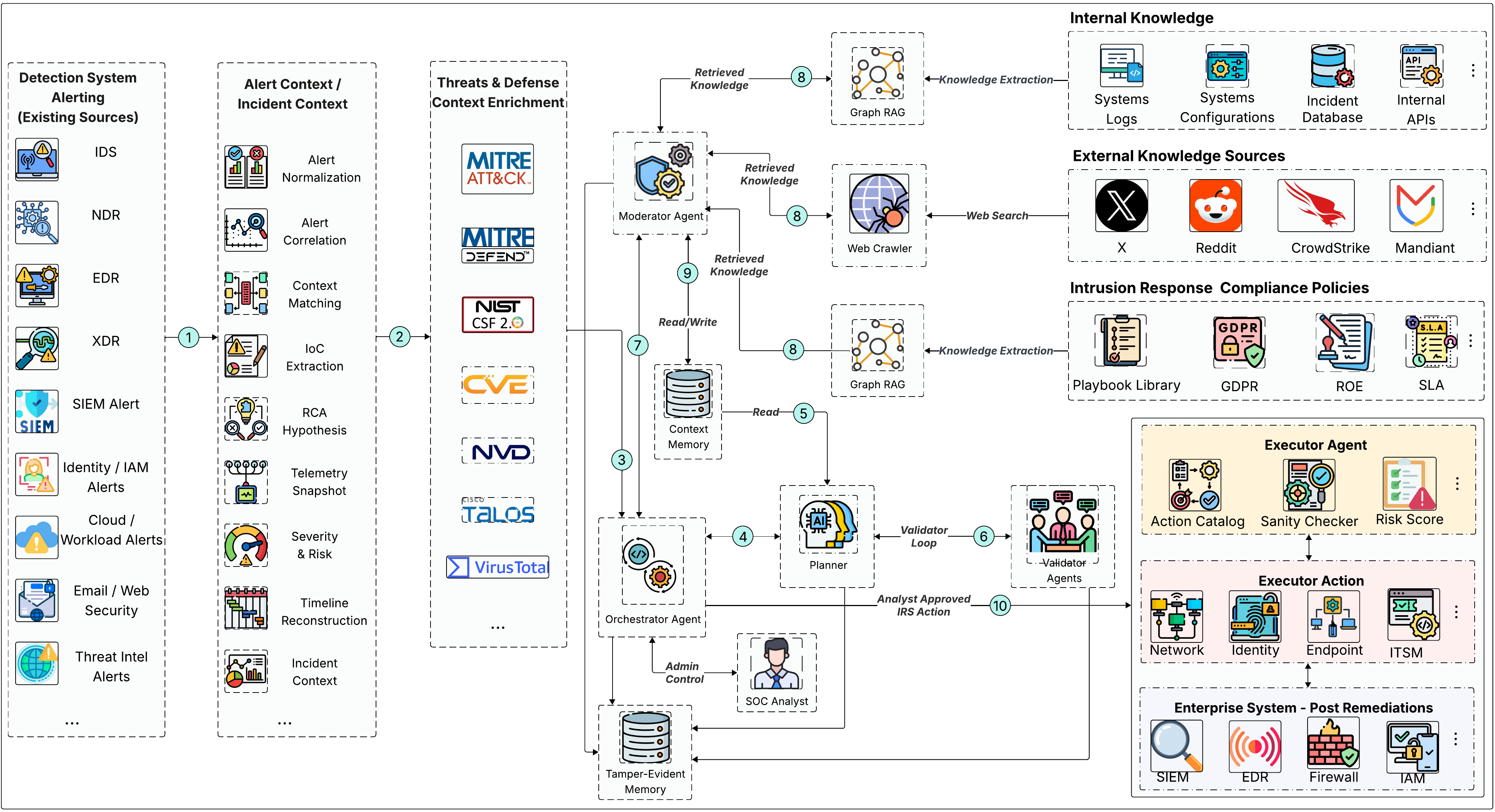}
\caption{The Agentra pipeline for agentic IRP generation and orchestration. Existing IDS, EDR, and XDR platforms supply alert evidence; Agentra normalizes this evidence into an Incident Context grounded in MITRE ATT\&CK, MITRE D3FEND, and NIST CSF~2.0 and enriches it with internal and external threat intelligence. A deterministic Orchestrator drives a bounded Planner--Validator review loop over Moderator-sanitized retrieval and, after analyst sign-off, executes through the Executor Agent's gated Action Catalog, with all activity recorded in a tamper-evident audit substrate.}
\label{fig:arch}
\end{figure*}

\subsection{Architectural Overview} \label{sec:overview}
Fig.~\ref{fig:arch} presents Agentra as a response-side pipeline that begins after primary detection. Existing detection platforms provide alert evidence, which Agentra normalizes into an Incident Context, enriches through moderated retrieval, and converts into a new or revised IRP. The IRP is the framework-level artifact; in our evaluation, it is serialized as an OASIS CACAO~v2.0 playbook, i.e., a Collaborative Automated Course of Action Operations security workflow~\cite{oasis_cacao_v2_2022}. Each IRP is validated, reviewed by an analyst, gated through execution controls, and logged.

The decomposition reflects scope control, independent review, and tool isolation. The Planner proposes candidate response plans but has no execution access. Validator agents review the plan independently, reducing reliance on a single model output. The Moderator mediates retrieval, so the Planner never consumes external threat intelligence directly. The Executor acts only through a bounded Action Catalog and risk gate. This separates reasoning, retrieval, validation, execution, and audit functions.

\subsection{Formal Orchestration Model} \label{sec:formal}
Algorithm~\ref{alg:agentra} formalizes the pipeline illustrated in Fig.~\ref{fig:arch}. Alert evidence $E$ is normalized into an enriched incident context $\mathcal{I}$. Internal Graph-RAG and external threat-intelligence retrievals produce $X_{\mathrm{int}}$ and $X_{\mathrm{ext}}$, which are screened by the Moderator before being exposed to the Planner as sanitized context $X$. To determine whether a response strategy requires adaptation, the Planner evaluates the existing playbook library $L$ against $\mathcal{I}$ and $X$. It then drafts a candidate incident response plan $P$, composed of discrete actions $a \in P$ that must later be checked against the Action Catalog $\mathcal{C}$.

Validator pool $\mathcal{V}=\{v_1,\ldots,v_n\}$ reviews the plan independently. Each Validator $v_i$ returns a binary approval vote $b_i \in \{0,1\}$ and a rationale $r_i$. The plan advances only if at least $k$ of the $n$ Validators approve it. The framework is specified for a general $k$-of-$n$ Validator pool; the reported evaluation instantiates this design with one Validator model per Planner--Validator combination under a bounded multi-round review loop. Larger heterogeneous Validator pools are therefore treated as extensions of the same orchestration model rather than evaluated claims in this study.

During execution, each proposed action is gated by a risk evaluation function $\mathrm{Risk}(a)=f(\kappa(a),\beta(a),\rho(a))$, where $\kappa(a)$ denotes asset criticality, $\beta(a)$ denotes blast radius, and $\rho(a)$ denotes reversibility. Policy thresholds $\Theta$ map this score to an authorization tier $\tau \in \{\textit{advise},\textit{approve},\textit{senior-approve}\}$. Through this structure, the Planner proposes, Validators review, the analyst authorizes, and the Executor acts only through predefined catalog and risk gates.

\begin{algorithm}[!ht]
\caption{Agentra supervisable IRP orchestration}
\label{alg:agentra}
\KwIn{Alert evidence $E$; playbook library $L$; Action Catalog $\mathcal{C}$; Validator pool $\mathcal{V}=\{v_1,\ldots,v_n\}$; approval threshold $k$; revision cap $R$; risk thresholds $\Theta$; append-only audit ledger $\mathcal{A}$}
\KwOut{Analyst-approved incident response plan $P^\star$, or terminal status: $\mathrm{escalate}$, $\mathrm{rejected}$, $\mathrm{invalidate}$, $\mathrm{deny}$, or $\mathrm{hold}$}

$\mathcal{I} \leftarrow \mathrm{BuildContext}(E)$;
$K \leftarrow \mathrm{ExtractKeys}(\mathcal{I})$;
$X_{\mathrm{int}} \leftarrow \mathrm{GraphRAG}(\mathcal{I},K)$;
$X_{\mathrm{ext}} \leftarrow \mathrm{WebRetrieve}(\mathcal{I},K)$;
$X \leftarrow \mathrm{Moderator}(X_{\mathrm{int}}\cup X_{\mathrm{ext}})$;
$\mathrm{Append}(\mathcal{A},\langle \text{enrichment},X,\mathrm{Prov}(X)\rangle)$;

$P \leftarrow \mathrm{Planner}(\mathcal{I},X,L)$;
$approved \leftarrow \mathrm{false}$;

\For{$t=1$ \KwTo $R$}{
$Y \leftarrow \{(\mathrm{vote}_i,\mathrm{rationale}_i)=v_i(P,\mathcal{I},X,\mathcal{C}) \mid v_i\in\mathcal{V}\}$;
$\mathrm{Append}(\mathcal{A},\langle \text{validator-votes},Y\rangle)$;
$approvals \leftarrow \sum_{(\mathrm{vote}_i,\mathrm{rationale}_i)\in Y} \mathrm{vote}_i$;

\If{$approvals \ge k$}{
    $approved \leftarrow \mathrm{true}$\;
    \textbf{break}\;
}

\If{$t<R$}{
    $P \leftarrow \mathrm{Planner}(\mathcal{I},X,P,Y)$\;
}
}

\If{$approved=\mathrm{false}$}{
$\mathrm{Escalate}(P,Y)$;
\Return{$\mathrm{escalate}$};
}

$d \leftarrow \mathrm{AnalystReview}(P)$;
$\mathrm{Append}(\mathcal{A},\langle \text{analyst-decision},d\rangle)$;

\If{$d\neq \mathrm{approve}$}{
\Return{$\mathrm{rejected}$};
}

\If{$\mathrm{StateChanged}(P)$}{
$\mathrm{Invalidate}(P)$;
\Return{$\mathrm{invalidate}$};
}

\ForEach{$a\in P$}{
\If{$a\notin\mathcal{C}$}{
\Return{$(\mathrm{deny},\mathrm{catalog})$};
}

\If{$\neg \mathrm{SanityOK}(a)$}{
    \Return{$(\mathrm{deny},\mathrm{consistency})$}\;
}

$s \leftarrow \mathrm{Risk}(a)$\;
$\tau \leftarrow \mathrm{Tier}(s,\Theta)$\;

\If{$\neg \mathrm{Authorized}(a,\tau)$}{
    \Return{$(\mathrm{hold},\tau)$}\;
}

$\mathrm{Execute}(a,\tau)$\;
$\mathrm{Append}(\mathcal{A},\langle \text{action},a,s,\tau\rangle)$\;
}

$P^\star \leftarrow P$;
\Return{$P^\star$};
\end{algorithm}

The remainder of this section explains the framework components represented in Algorithm~\ref{alg:agentra}: Incident Context construction, Validator-based IRP generation, moderated knowledge augmentation, gated execution, and auditability.

\subsection{Incident Context Construction} \label{sec:context}
Incident Context Construction converts heterogeneous alert evidence into a compact typed record for response planning. The record includes IoCs, root-cause hypothesis, telemetry snapshot, severity, timeline, ATT\&CK technique, D3FEND mitigation candidates, and NIST CSF function. Where upstream tools have not supplied these mappings, Agentra enriches the record through CVE, NVD, Cisco Talos, VirusTotal, MITRE ATT\&CK, MITRE D3FEND~\cite{kaloroumakis_d3fend_2021}, and NIST CSF~2.0~\cite{nist_csf2_2024}. This typed interface prevents response agents from reading raw telemetry directly and preserves modularity between detection and response.

\subsection{Supervisable IRP Generation and Validation} \label{sec:orchestration}
This stage implements Agentra's core supervisability mechanism. The deterministic Orchestrator coordinates agents but does not decide the response. After reading Moderator-sanitized context, the \textbf{Planner} drafts a plan composed of discrete candidate actions. The Planner has no execution access, so proposed actions must later pass catalog, sanity, and risk gates.

The candidate plan, reasoning trace, and supporting context are sent to $n$ \textbf{Validator Agents}. Each Validator $v_i$ evaluates the plan against the Incident Context, sanitized retrieved context, compliance-policy library, and Action Catalog $\mathcal{C}$, returning a binary approval vote $\mathrm{vote}_i\in{0,1}$ and rationale $\mathrm{rationale}_i$. In the general framework, the plan advances only if at least $k$ of the $n$ Validators approve it. In the evaluated configuration, this mechanism is instantiated as a bounded Planner--Validator review loop with one Validator per model combination. Rejected candidates return to the Planner with structured feedback for a bounded number of revisions; if approval is not reached, the case is escalated to the SOC with the Validator transcript. We use Byzantine-style terminology as an analogy, while recognizing that shared base models do not satisfy classical independent-failure assumptions. Agentra approximates independence through role-diverse prompts, diverse backends where feasible, and diverse decoding parameters.

\subsection{Knowledge Augmentation through a Moderated Security Gateway} \label{sec:moderatedrag}
Knowledge augmentation is mediated by a dedicated \textbf{Moderator agent}, so the Planner never retrieves directly. The Moderator uses retrieval keys $K$ from the Incident Context $\mathcal{I}$ to query two trust tiers. Internally, it retrieves logs, configurations, incident records, internal API data, IRP library entries, and compliance policies. This internal retrieval is served through Graph-RAG, which captures relationships among logs, assets, incidents, and policies~\cite{edge2024graphRAG}. Externally, the Moderator performs controlled retrieval over OSINT and threat-intelligence sources.

External content remains untrusted. Prior work demonstrates indirect prompt injection through retrieved content~\cite{greshake2023indirectInjection} and vulnerabilities in tool-using agents~\cite{debenedetti2024agentdojo}. The Moderator therefore screens retrieved material for embedded instructions, indirect injection, allow-list violations, out-of-ontology hallucinated mitigations, contradictions with internal graph evidence, cross-site scripting, man-in-the-middle interference, data poisoning, and tool poisoning. Suspect content is rewritten with provenance metadata or rejected. The Moderator has no execution authority and writes only sanitized context to Context Memory. In this paper, the Moderator is evaluated as a designed control within the response pipeline; we do not claim that its prompt-injection robustness has been fully validated against an adaptive adversarial retrieval suite.

\subsection{Gated Execution and Auditability} \label{sec:gated}
Analyst sign-off authorizes response strategy, not unmediated execution. Each action passes through three gates. The \textbf{Action Catalog} defines permitted actions across Network, Identity, Endpoint, and ITSM surfaces, including host isolation, token revocation, IoC blocking, and ticket creation; actions outside this catalog cannot execute, addressing the IRP-to-API mapping problem identified by Sworna et al.~\cite{sworna2023IRP2APIplanMapping}. The \textbf{Sanity Checker} verifies internal consistency, including dependencies on assets already quarantined by earlier actions. The \textbf{Risk Score} combines asset criticality $\kappa(a)$, blast radius $\beta(a)$, and reversibility $\rho(a)$ into the authorization tiers \emph{advise}, \emph{approve}, and \emph{senior approve}; in the evaluation harness, these values are discretized from the static service-dependency overlay and mapped using fixed thresholds $\Theta$. If the asset, dependency, or remediation state required by $P$ differs from the state snapshot used during planning, $\mathrm{StateChanged}(P)$ invalidates the plan and triggers reanalysis.

Agentra records prompts, retrievals, Moderator decisions, Validator votes, Risk Score outcomes, Sanity Checker outcomes, analyst decisions, and executed actions in a tamper-evident append-only audit substrate. Unlike mutable Context Memory, this substrate is write-once: records are hashed as Merkle-tree leaves, and each audit batch stores a Merkle root chained to the prior root, making deletion, insertion, or retroactive modification detectable. These controls are complementary: the Moderator reduces poisoned retrieval, the Validator review loop reduces single-planner failure, execution gates constrain action scope, and the audit log preserves review evidence.

\section{Experimental Evaluation}\label{sec:experiments}
Evaluating an IRS requires metrics beyond playbook emission. A response system must identify when an alert warrants action, generate a grounded plan, avoid excessive disruption, and remain usable under incident-response latency constraints. We evaluate Agentra using four main measures: FP-aware IRS $F_{1}$, false-positive rate, D3FEND coverage and precision, and projected harmful-action rate; latency is reported as a batch-triage observation. The ablation adds Agentra components over a static CACAO baseline: A0 is CACAO lookup only, A1 adds the Planner, A2 adds Validator debate, A3 adds moderated threat-intelligence retrieval, and A4 adds Graph-RAG enterprise grounding.

\subsection{Evaluation Scope and Corpus}
The evaluation uses a 120-event corpus drawn from ThreatHunter-Playbook~\cite{rodriguez_threathunter_playbook}, Splunk BOTSv3~\cite{splunk_botsv3_2018}, and DARPA OpTC~\cite{darpa_optc_2020}. The corpus spans malicious intrusion activity, benign administrative activity, and false-positive stress cases in which benign event content is paired with an incorrect malicious label. This design tests both response coverage and misattribution robustness. The final split is $58$ malicious events and $62$ benign or false-positive stress events. The malicious events cover 43 MITRE ATT\&CK techniques/sub-techniques across credential access, discovery, execution, persistence, privilege escalation, defense evasion, command and control, exfiltration, and impact. Each event is deterministically anonymized before LLM processing by replacing host names, account identifiers, emails, public IPs, domains, and access keys with stable placeholders or documentation-range addresses.

To preserve reproducibility, all ontology and framework artifacts used for grounding and scoring were cached at evaluation time, including the MITRE ATT\&CK, MITRE D3FEND, NIST CSF~2.0, and CACAO references used by the harness. The evaluation therefore does not depend on live changes to online ontology pages or revised framework documentation after the experiment date.

Listing~\ref{lst:event} shows representative corpus entries. External threat-dataset scenarios are restaged as Kubernetes audit events before LLM processing. This restaging provides a common event format while preserving the ATT\&CK technique, event summary, and benign or malicious ground truth used by the evaluation harness.

\begin{lstlisting}[style=ag,caption={Representative corpus events. External threat-dataset scenarios are re-staged as Kubernetes audit events; identifiers are anonymized before any LLM call.},label={lst:event}]
# THP - Credential Access (T1003.001): Mimikatz LSASS read, re-staged
# as kubectl exec into a pod by a non-approved service account.
* id: EVT-THP-001
  dataset: THP
  ttp: T1003.001
  ground_truth: malicious
  raw_summary: "Mimikatz sekurlsa::logonpasswords -- process opens handle to LSASS"
  state: {audit_event: {verb: create, subresource: exec, resource: pods,
  actor: dev-user-sa}}

# BOTS - Persistence (T1098.001): compromised IAM identity requests
# a new credential; k8s analogue is a TokenRequest.
* id: EVT-BOTS-001
  dataset: BOTS
  ttp: T1098.001
  ground_truth: malicious
  raw_summary: "Compromised IAM identity attempted to create a new access key"
  state: {audit_event: {verb: create, resource: serviceaccounts/token,
  actor: anonymous-or-stolen-sa, source_ip: 203.0.113.10}}

# OpTC - Discovery (T1087.002): no CACAO rule exists for this TTP.
* id: EVT-OPTC-016
  dataset: OPTC
  ttp: T1087.002
  ground_truth: malicious
  raw_summary: "Bulk Domain Admin enumeration"
  state: {}

# Benign control later used in a false-positive stress case.
* id: EVT-BOTS-030
  dataset: BOTS
  ttp: null
  ground_truth: benign
  state: {dns_resolution_failures_per_minute: 1}
  \end{lstlisting}

Reference D3FEND labels are curated by mapping each malicious event's ATT\&CK technique to the D3FEND~v1.0 ontology over a closed 44-identifier vocabulary. We treat these as reference sets rather than exhaustive ground truth because the mapping depends on analyst judgment and ontology interpretation. The TTP-to-state translation is deterministic and auditable, so each mapping is reproducible and independently re-derivable. Because all datasets pre-date the evaluated models and are widely indexed online, anonymization reduces but does not eliminate dataset-recognition risk.

\subsection{Metrics and Statistical Protocol} \label{sec:metric-defs}
For each event $e$, let $C_e$ be the set of D3FEND identifiers cited by the emitted playbook and $R_e$ be the curated reference set. An event is scored as response-positive when $C_e\cap R_e\neq\emptyset$ and response-negative otherwise. We report precision, recall, and $F_1=2PR/(P+R)$ over the 120-event corpus, along with the false-positive rate on stress cases. Response quality is measured using D3FEND coverage, $\mathrm{Cov}_e=|C_e\cap R_e|/|R_e|$, and D3FEND precision, $\mathrm{Prec}^{D3}_e=|C_e\cap R_e|/|C_e|$. Projected safety is measured by the harmful-action rate: a playbook is flagged harmful if it recommends an irreversible action for a reconnaissance or discovery-stage event. This conservative analytical overlay is not a live execution measurement or comprehensive safety guarantee.

All results are reported for a single seed (seed~42); multi-seed dispersion and per-seed McNemar significance testing are deferred to the extended version. We therefore report point estimates and emphasize the direction and magnitude of within-combination ablation deltas rather than formal significance. Listing~\ref{lst:playbook} and Listing~\ref{lst:d3fend} illustrate how an emitted CACAO playbook is scored against the curated D3FEND reference set.

\begin{lstlisting}[style=ag,caption={Emitted CACAO v2.0 response playbook (Agentra A4) for EVT-OPTC-016 after a 2-round Planner--Validator debate. Each step cites a D3FEND countermeasure id.},label={lst:playbook}]
{ "type": "playbook", "playbook_types": ["T1087.002"],
"workflow": [
{"name":"enable-account-monitoring",
"agent":"identity",
"d3fend_ref":["D3-AM"],
"command":"Enhanced account monitoring with LDAP activity baselining."},
{"name":"analyze-authentication-logs",
"agent":"identity",
"d3fend_ref":["D3-ANAA"],
"command":"Cross-reference auth logs with domain-account enumeration."},
{"name":"monitor-endpoint-activity",
"agent":"endpoint",
"d3fend_ref":["D3-OSM"],
"command":"OS-level monitoring for anomalous LDAP client behavior."} ] }
\end{lstlisting}

\begin{lstlisting}[style=ag,caption={ATT\&CK-to-D3FEND reference mapping and scoring for the playbook in Listing~\ref{lst:playbook}.},label={lst:d3fend}]
# Reference D3FEND set for the event's TTP:
T1087.002 -> {D3-AM   (Account Monitoring,          detect/identity),
D3-ANAA (Auth. Activity Analysis,     detect/identity),
D3-OSM  (Operating System Monitoring, detect/endpoint) }

# Scoring the emitted playbook:
# cited     = {D3-AM, D3-ANAA, D3-OSM}
# reference = {D3-AM, D3-ANAA, D3-OSM}
# coverage  = |cited n reference| / |reference| = 3/3 = 1.00
# precision = |cited n reference| / |cited|     = 3/3 = 1.00
\end{lstlisting}

\subsection{Ablation Configurations and Model Backends} \label{sec:ablations}
The additive ablation builds Agentra one component at a time over the CACAO baseline (Table~\ref{tab:ablations}).

\begin{table}[t]
\centering
\caption{Additive ablations. A4 is the full Agentra stack.}
\label{tab:ablations}
\scriptsize
\begin{tabular}{@{}p{0.10\columnwidth}p{0.84\columnwidth}@{}}
\toprule
ID & Definition \\
\midrule
A0 & CACAO lookup only; no LLM. \\
A1 & A0 + Planner: keeps, extends, or synthesizes a CACAO playbook. \\
A2 & A1 + Validator debate. \\
A3 & A2 + Web-crawler + Moderator: local CTI grounding, sanitization, D3FEND breadth check. \\
A4 & A3 + Graph-RAG: enterprise-config grounding; full stack. \\
\bottomrule
\end{tabular}
\end{table}

We evaluate five Planner/Validator combinations under the same protocol (Table~\ref{tab:model-combos}). M1 uses the \emph{replace-CACAO} architecture, in which the Planner synthesizes from scratch; M2--M5 use \emph{enhance-CACAO}, in which the catalog entry seeds the Planner. Local models are served through vLLM with FP8 quantization on a single NVIDIA~H200~NVL with $143\,\mathrm{GB}$ HBM3e. Foundation-Sec-8B, Qwen3-32B, and Mistral-Small-3.2-24B are used at their public releases~\cite{foundationsec_2025,qwen3_2025,mistral_small_2025}.

The evaluation uses representative Planner--Validator pairs rather than simultaneous multi-validator quorums. Under the context length and serving configuration used in this study, a single NVIDIA H200 NVL could co-host the Planner and one Validator from the selected model class, but not several large heterogeneous Validators in parallel. Sequential multi-validator quorums are compatible with the framework, but would change the latency profile and were outside the compute envelope of this submission. We therefore report the single-Validator instantiation as a controlled evaluation of the Planner--Validator review loop and leave larger $k$-of-$n$ deployments to future work.

\begin{table}[t]
\centering
\caption{Planner and Validator combinations.}
\label{tab:model-combos}
\footnotesize
\begin{tabular}{@{}llll@{}}
\toprule
ID & Planner & Validator & Mode \\
\midrule
M1 & Qwen3-32B-Instr. & Mistral-Small-3.2-24B & replace-CACAO \\
M2 & Qwen3-32B-Instr. & Mistral-Small-3.2-24B & enhance-CACAO \\
M3 & Qwen3-32B-Instr. & Foundation-Sec-8B-Reas. & enhance-CACAO \\
M4 & Claude Sonnet 4.6 & Foundation-Sec-8B-Reas. & enhance-CACAO \\
M5 & GPT-5 & Claude Sonnet 4.6 & enhance-CACAO \\
\bottomrule
\end{tabular}
\end{table}

\subsection{Results and Analysis} \label{sec:rq2}
Table~\ref{tab:headline} reports the full FP-aware IRS results. The static CACAO baseline (A0, identical across combinations) is an ATT\&CK-keyed CACAO catalog mapped to the same closed D3FEND vocabulary as Agentra. It matches $49/120$ events and attains recall $0.55$ and $F_1{=}0.61$, but fires on half of the false-positive stress cases ($\mathrm{FP\text{-}rate}{=}0.50$) because lookup uses the supplied TTP rather than benign-content features. The emission ceiling ($49$) exceeds the $\mathrm{TP}{+}\mathrm{FP}$ count ($47$) in Table~\ref{tab:headline} because two emitted catalog playbooks had no D3FEND identifier overlapping the event reference set and were therefore scored as response-negative.

\paragraph{A1 (Planner): large recall gain, planner-dependent false positives}
Adding the Planner raises response-decision recall for capable back-ends (M4 $0.55{\to}0.90$; M5 $0.91$), but the output is operationally raw: D3FEND precision collapses ($\mathrm{Prec}^{D3}{\le}0.30$ for every combination) and the harmful-action rate rises to $14$--$16\%$ on the API planners. False-positive behavior splits by planner: local planners suppress misattributions more often, whereas M4 emits a playbook for almost every misattributed event ($\mathrm{FP\text{-}rate}{=}0.97$). More permissive planners are therefore more vulnerable to TTP misattribution.

\paragraph{A2 (Validator): combination-dependent}
Bounded debate sharpens some combinations and destabilizes others. It gives M1 its lowest false-positive rate ($0.20$) but does not lift response quality, and on M5 it collapses recall to $0.14$: the strict Sonnet validator rejects most GPT-5 proposals, and many events exhaust the $600\,\mathrm{s}$ debate budget. Without the closed-vocabulary and breadth checks added at A3, debate alone cannot reliably separate useful corrections from spurious citations.

\paragraph{A3 (Moderator + CTI): the response-quality transition}
Adding the Web-crawler and Moderator is the qualitative inflection across all combinations: D3FEND precision jumps (M1 $0.23{\to}0.79$; M4 $0.15{\to}0.68$; M5 $0.20{\to}0.95$) and the harmful-action rate falls to at most $0.9\%$, while recall recovers for the local combinations. The Moderator's sanitization and post-hoc D3FEND breadth check narrow the citation set, and CTI grounding helps the Planner drop risky actions such as cluster-wide isolation or irreversible steps on reconnaissance-only TTPs.

\paragraph{A4 (full stack) and the recall--precision tradeoff}
Graph-RAG enterprise grounding completes the stack and yields the best results, but the FP-aware view reveals a tradeoff hidden by malicious-only evaluation. M4 attains perfect recall ($58/58$) yet the weakest false-positive control ($\mathrm{FP\text{-}rate}{=}0.90$), so its precision ($0.68$) caps $F_1$ at $0.81$. M1 balances strong recall ($0.90$) with moderate false positives ($\mathrm{FP\text{-}rate}{=}0.47$), achieving the highest FP-aware IRS $F_1$ ($0.84$). M5 produces the cleanest playbooks (D3FEND precision $0.95$) but at lower recall ($0.57$). No single combination dominates: capability maximizes recall, but grounding and validation layers govern false-positive robustness and response quality.

\subsection{Cross-Combination Behavior} \label{sec:rq3}
\begin{figure*}[t]
\centering
\includegraphics[width=\textwidth,height=2.1in]{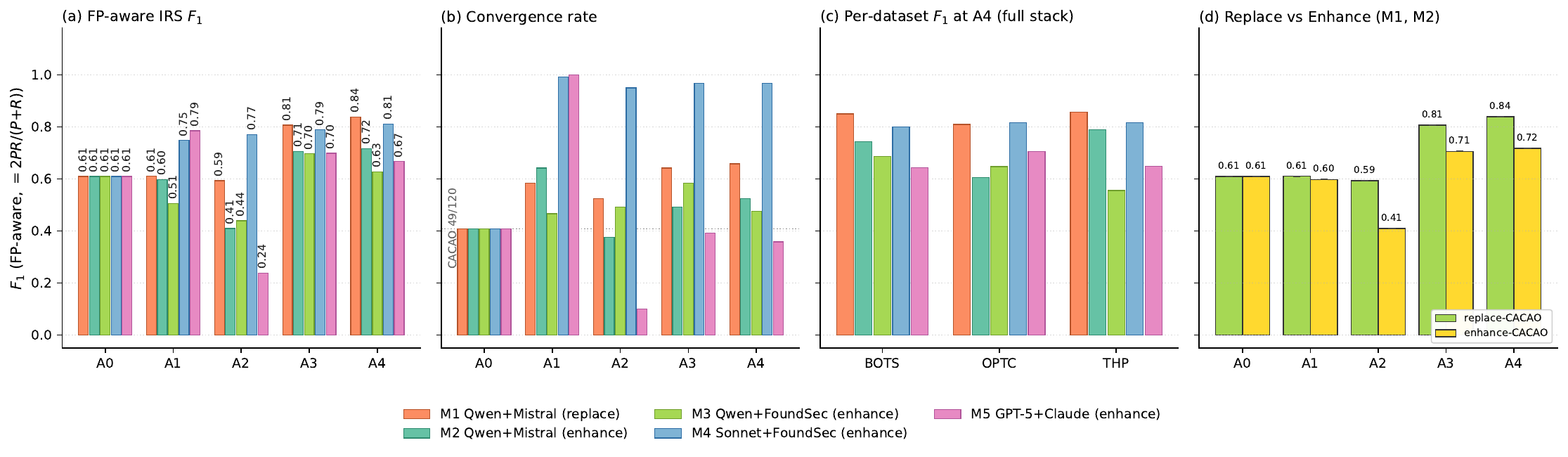}
\caption{Cross-combination results under the additive ablation ($120$-event FP-aware corpus, seed 42). \textbf{(a)} FP-aware IRS $F_1{=}2PR/(P{+}R)$ by ablation. \textbf{(b)} Convergence rate, i.e., Validator-approved plans within the revision/latency budget; the dotted line marks the CACAO emission ceiling ($49/120$). \textbf{(c)} Per-dataset $F_1$ at A4. \textbf{(d)} Replace-CACAO (M1) versus enhance-CACAO (M2). M4 attains the highest recall but the weakest false-positive control; M1 attains the highest FP-aware IRS $F_1$.}
\label{fig:main}
\end{figure*}

Figure~\ref{fig:main} compares the five combinations on FP-aware IRS $F_1$, convergence rate, per-dataset $F_1$ at A4, and the replace- versus enhance-CACAO pair (M1/M2). Two patterns hold. First, the additive ladder improves $F_1$ and response quality across every combination, with A3 the consistent safety and quality inflection. Second, false-positive robustness is weaker for more permissive planners: the local-planner combinations (M1--M3) hold $\mathrm{FP\text{-}rate}\le0.47$ at A4, whereas API-planner combinations over-emit on misattributed events. Replace-CACAO (M1) edges enhance-CACAO (M2) at A4 ($0.84$ versus $0.72$), indicating that for a strong local planner the grounded A3 and A4 stages compensate for the absent catalog starting point.

Finally, THP, BOTS, and OpTC pre-date all evaluated model training cutoffs and are widely indexed online, so residual memorization remains a validity threat that anonymization reduces but cannot eliminate. We therefore weight within-combination comparisons above absolute cross-model rankings. End-to-end latency on the M4 stack averages approximately $40\,\mathrm{s}$ per event, dominated by the Planner--Validator debate, and is reported as a batch-triage figure rather than a real-time guarantee.

\newcommand{\wincell}[1]{\cellcolor{winrow}#1}

\begin{table}[t]
\centering
\caption{FP-aware results on the $120$-event corpus (seed 42). Counts are TP/FN and FP/TN; $P$, $R$, and $F_1$ are response-decision precision, recall, and their harmonic mean; FP$_{\mathrm{stress}}$ is the false-positive rate on stress cases; Cov.\ and Prec.\ are mean D3FEND coverage and precision; Harm\% is projected harmful-action rate. Rates: 2 decimals; Harm\%: 1 decimal. Blue rows mark per-ablation $F_1$ winners; green marks best recall; orange marks best D3FEND coverage/precision.}
\label{tab:headline}
\footnotesize
\setlength{\tabcolsep}{1pt}
\begin{tabular}{ll|cc|ccc|c|cc|c}
\toprule
& & \multicolumn{2}{c|}{Counts} & \multicolumn{3}{c|}{Response} & \multicolumn{1}{c|}{Stress} & \multicolumn{2}{c|}{D3FEND} & \multicolumn{1}{c}{Safety} \\
\cmidrule(lr){3-4}\cmidrule(lr){5-7}\cmidrule(lr){9-10}
Combo & Abl. & TP/FN & FP/TN & $P$ & $R$ & $F_1$ & FP$_{\mathrm{stress}}$ & Cov. & Prec. & Harm\% \\
\midrule

\multirow{5}{*}{M1} 
& A0 & 32/26 & 15/47 & 0.68 & 0.55 & 0.61 & 0.50 & 0.37 & 0.85 & 0.0 \\
& A1 & 29/29 & 8/54 & 0.78 & 0.50 & 0.61 & 0.27 & 0.16 & 0.22 & 4.3 \\
& A2 & 27/31 & 6/56 & 0.82 & 0.47 & 0.59 & 0.20 & 0.17 & 0.23 & 6.3 \\
& \wincell{A3} & \wincell{48/10} & \wincell{13/49} & \wincell{0.79} & \wincell{0.83} & \wincell{0.81} & \wincell{0.43} & \wincell{0.79} & \wincell{0.79} & \wincell{0.0} \\
& \wincell{A4} & \wincell{52/6} & \wincell{14/48} & \wincell{0.79} & \wincell{0.90} & \wincell{0.84} & \wincell{0.47} & \wincell{0.84} & \wincell{0.84} & \wincell{0.0} \\
\midrule

\multirow{5}{*}{M2} 
& A0 & 32/26 & 15/47 & 0.68 & 0.55 & 0.61 & 0.50 & 0.37 & 0.85 & 0.0 \\
& A1 & 32/26 & 17/45 & 0.65 & 0.55 & 0.60 & 0.57 & 0.24 & 0.30 & 10.4 \\
& A2 & 17/41 & 8/54 & 0.68 & 0.29 & 0.41 & 0.27 & 0.19 & 0.26 & 2.2 \\
& A3 & 36/22 & 8/54 & 0.82 & 0.62 & 0.71 & 0.27 & 0.75 & 0.75 & 0.0 \\
& A4 & 38/20 & 10/52 & 0.79 & 0.66 & 0.72 & 0.33 & 0.76 & 0.76 & 0.0 \\
\midrule

\multirow{5}{*}{M3} 
& A0 & 32/26 & 15/47 & 0.68 & 0.55 & 0.61 & 0.50 & 0.37 & 0.85 & 0.0 \\
& A1 & 23/35 & 10/52 & 0.70 & 0.40 & 0.51 & 0.33 & 0.24 & 0.30 & 5.4 \\
& A2 & 20/38 & 13/49 & 0.61 & 0.34 & 0.44 & 0.43 & 0.23 & 0.27 & 8.5 \\
& A3 & 38/20 & 13/49 & 0.75 & 0.66 & 0.70 & 0.43 & 0.72 & 0.73 & 0.0 \\
& A4 & 32/26 & 12/50 & 0.73 & 0.55 & 0.63 & 0.40 & 0.77 & 0.77 & 0.0 \\
\midrule

\multirow{5}{*}{M4} 
& A0 & 32/26 & 15/47 & 0.68 & 0.55 & 0.61 & 0.50 & 0.37 & 0.85 & 0.0 \\
& A1 & 52/6 & 29/33 & 0.64 & 0.90 & 0.75 & 0.97 & 0.37 & 0.15 & 16.0 \\
& \wincell{A2} & \wincell{52/6} & \wincell{25/37} & \wincell{0.68} & \wincell{0.90} & \wincell{0.77} & \wincell{0.83} & \wincell{0.37} & \wincell{0.15} & \wincell{17.5} \\
& A3 & 56/2 & 28/34 & 0.67 & 0.97 & 0.79 & 0.93 & 0.72 & 0.68 & 0.9 \\
& A4 & 58/0 & 27/35 & 0.68 & \cellcolor{recallc}1.00 & 0.81 & 0.90 & 0.73 & 0.68 & 0.0 \\
\midrule

\multirow{5}{*}{M5} 
& A0 & 32/26 & 15/47 & 0.68 & 0.55 & 0.61 & 0.50 & 0.37 & 0.85 & 0.0 \\
& \wincell{A1} & \wincell{53/5} & \wincell{24/38} & \wincell{0.69} & \wincell{0.91} & \wincell{0.79} & \wincell{0.80} & \wincell{0.31} & \wincell{0.23} & \wincell{14.2} \\
& A2 & 8/50 & 1/61 & 0.89 & 0.14 & 0.24 & 0.03 & 0.42 & 0.20 & 33.3 \\
& A3 & 36/22 & 9/53 & 0.80 & 0.62 & 0.70 & 0.30 & \cellcolor{qualc}0.96 & \cellcolor{qualc}0.95 & 0.0 \\
& A4 & 33/25 & 8/54 & 0.81 & 0.57 & 0.67 & 0.27 & \cellcolor{qualc}0.95 & \cellcolor{qualc}0.95 & 0.0 \\
\bottomrule
\end{tabular}
\end{table}

\section{Discussion and Limitations}\label{sec:discussion}
Agentra is strongest where its claims are architectural and most limited where they depend on model behavior. The Validator review loop, Moderator gateway, Action Catalog, tiered Risk Score, and append-only audit substrate are structural controls that remain part of the framework regardless of the model assigned to each role. By contrast, response quality depends on emergent behaviors such as Planner over-generation, Validator sensitivity, timeout patterns, and whether multiple agents provide independent reasoning. The evaluation therefore supports Agentra as a model-agnostic architecture but not as a model-invariant system.

The reported experiments instantiate the general Validator-pool design with one Validator per Planner--Validator combination. This limits claims about Validator diversity: the results evaluate bounded Planner--Validator review, not the full fault-tolerance benefits of a heterogeneous $k$-of-$n$ Validator pool. Under the serving configuration used in this study, a single NVIDIA H200 NVL could host the Planner and one Validator from the selected model class, but not several large Validators in parallel. We plan to evaluate larger heterogeneous Validator pools using additional accelerators, which will also require a separate latency study.

The additive ablation clarifies the role of the architectural stages. On the M4 reference stack, A1 closes most of the CACAO-to-recall gap, but with low D3FEND precision ($0.15$) and a projected harmful-action rate of $16.0\%$. A2 does not improve this trade-off on the evaluated corpus. The qualitative transition occurs at A3, where threat-intelligence grounding and Moderator control raise D3FEND precision to $0.68$ and reduce the projected harmful-action rate to $0.9\%$. On M4, A4 adds Graph-RAG enterprise grounding, attains perfect recall, and removes the residual projected harmful actions ($0.0\%$). This harmful-action result should be interpreted relative to the ablation path: the static CACAO baseline already has a projected harmful-action rate of $0.0\%$, Planner-only configurations introduce unsafe overreaction, and the A3/A4 grounding layers remove those Planner-induced harmful actions while preserving much of the recall gain.

The evaluation does not include an adversarial retrieval benchmark for the Moderator. Static corpora can support prompt-injection evaluation by planting malicious instructions in fixed retrieved documents, as prior work has shown; therefore, this is an evaluation-scope limitation rather than a limitation imposed by the corpus format. In this paper, Moderator screening is evaluated as part of the designed response pipeline, while adaptive prompt-injection and poisoned-retrieval robustness are deferred to future work.

Several additional limitations bound the results. The corpus is controlled and based on public datasets, so deterministic anonymization reduces but does not eliminate memorization risk. D3FEND labels are curated reference sets rather than exhaustive ground truth, and the study reports single-seed point estimates rather than multi-seed confidence intervals. Operational safety is projected using a static dependency overlay rather than measured in deployment, and analyst approval is simulated in the harness. These constraints limit claims about production effectiveness, but they do not affect the within-ablation comparison of Agentra components under the same evaluation protocol. In future work, we plan to evaluate Agentra in a simulated enterprise environment or cyber range to measure MTTD, MTTR, playbook effectiveness, observed harmfulness, sanitization robustness, analyst burden, and response blast radius.

\section{Conclusion and Future Work}\label{sec:conclusion}
Agentra frames intrusion response as supervisable multi-agent planning rather than direct autonomous execution. It converts alerts from existing detection platforms into ontology-grounded CACAO plans, validates proposed plans through a bounded Planner--Validator review loop, screens retrieved knowledge through a Moderator gateway, gates actions through catalog and risk checks, and records decisions in an append-only audit substrate. On a 120-event corpus, the full Agentra stack improves FP-aware IRS $F_1$ from $0.61$ to $0.84$ over the static CACAO baseline and restores the projected harmful-action rate to the static baseline level of $0.0\%$ after Planner-only configurations introduce unsafe overreaction. These results support Agentra as a response-planning architecture.

In future work, we plan to evaluate Agentra in a simulated enterprise environment or cyber range rather than only on a static corpus. Such an environment would allow direct measurement of mean time to detect (MTTD), mean time to respond (MTTR), playbook completion rate, response effectiveness, operational false positives, observed blast radius, analyst approval burden, and the gap between projected and realized harmfulness. We also plan adversarial retrieval experiments in which a red-team LLM or adversarial content generator injects malicious instructions into retrieved threat intelligence, incident notes, or internal documents to test Moderator robustness under controlled attack conditions.

\section{Acknowledgment}
    The authors acknowledge the use of Flaticon (\url{www.flaticon.com}) in paper's figures. Any opinions, findings, conclusions, or recommendations expressed in this material are those of the authors and do not necessarily reflect the views of their institution or the funding agencies.

\bibliographystyle{ieeetr}
\bibliography{sources}

\end{document}